# Probing moiré electronic structures through quasiparticle interference

Luke C. Rhodes[1,*], Dylan C. Houston[1,†], Olivia R. Armitage[1], and Peter Wahl[1,2]

[1]*SUPA, School of Physics and Astronomy, University of St Andrews, North Haugh, St Andrews, Fife, KY16 9SS, United Kingdom*
[2]*Physikalisches Institut, Universität Bonn, Nussallee 12, 53115 Bonn, Germany*

Moiré lattices are a general feature of bilayer structures, where an additional periodic superstructure is generated by either lattice mismatch or from a twist angle. They have been shown to stabilize exotic ground states, including unconventional superconductivity and Mott insulating phases, attributed to strong electron correlations. However, controlling these interactions requires a detailed understanding of the low-energy electronic structure, which is lacking so far. Probing the electronic structure is challenging due to sample inhomogeneity, the low characteristic energy scales involved, and small sample sizes. Through quasiparticle interference (QPI) imaging, scanning tunneling microscopy (STM) can overcome many of these challenges but requires detailed modeling to extract the $k$-space electronic structure. Here, we present realistic calculations of QPI in twisted bilayer structures, which accounts for the effect of the long-range moiré lattice on the electronic structure as well as its interaction at a defect. These calculations reveal that, while the moiré supercell significantly reduces the size of the Brillouin zone, the QPI scattering vectors retain characteristics of the individual monolayers with distinct perturbations from the twisted geometry that can be directly linked back to the electronic structure. The procedure introduced here provides a general framework to use QPI to determine the low-energy electronic structure in moiré lattice systems.

The discovery of correlated phases and unconventional superconductivity in twisted bilayer graphene (TBLG) [1,2] established moiré lattices as a materials platform to study and control correlated ground states [2–6]. Stacking two graphene layers misaligned by some angle creates a new periodicity that enlarges the unit cell, resulting in backfolding effects, Van Hove singularities (VHs), and the formation of flat bands at a magic angle [7–9].

The control of the electronic states afforded by the twist angle is not specific to TBLG but general to any bilayer structure with a lattice mismatch or misalignment. Similar correlated ground states have been explored in a growing number of bilayer systems [10–13]. Understanding how these correlated states emerge as well as their nature requires a detailed knowledge of the low-energy electronic structure, which is, however, difficult to obtain.

Spectroscopic characterization of the moiré electronic structure is hampered by the small device sizes coupled with inhomogeneity of the twist angle as well as the small energies at which correlated behavior emerges, limiting the experimental techniques that can be used. Quasiparticle interference (QPI) imaging, measured via scanning tunneling microscopy (STM), can overcome these issues by locally measuring the modulations in the local density of states induced by defects with submillielectronvolt energy resolution and at millikelvin temperatures. However, two challenges arise: (1) QPI provides indirect information about the band structure in the form of scattering vectors $\mathbf{q} = \mathbf{k} - \mathbf{k}'$, rendering the extraction of the energy-momentum dispersion $E(\mathbf{k})$ nontrivial. (2) The moiré lattice generates exceptionally large unit cells, which suggests that large areas need to be sampled in real space to achieve sufficiently high $\mathbf{q}$-space resolution, creating experimental challenges.

Simulations of QPI can mitigate these challenges, although they come with their own difficulties, particularly in systems with large numbers of atoms within the unit cell. Most current theoretical work balances between high intraunit-cell resolution in a small field of view [14] and large fields of view with little to no intraunit-cell resolution [15–18]. The former approach is ideal to study the shape of localized defects but inefficient at simulating larger areas required to obtain Fourier transformations of QPI with high resolution. The latter approach overcomes this but at the expense of neglecting critical details of the tunneling matrix elements required for quantitative comparison between theory and experiment, such as how the wave functions of the sample overlap with the STM tip and how the electronic structure of the moiré system shows up in the real-space local density of states.

In this letter, we overcome these challenges by employing a continuum Green's function technique previously applied to iron-based superconductors, cuprates, and ruthenate systems [19–23]. This approach enables the calculation of the continuum local density of states (cLDOS) in real space, incorporating the effects of the long-range moiré lattice as well as local QPI from defects. Based on this formalism, we make realistic predictions for how the moiré lattice will influence QPI signatures in Fourier space for TBLG, which can be tested by experimental measurements. We can directly connect the intensity and magnitude of the scattering vectors to the electronic structure via comparison with the unfolded spectral function [24,25], providing a route to experimentally probe the low-energy electronic structure of moiré systems from measurements of QPI.

*Contact author: lcr23@st-andrews.ac.uk
†Present address: School of Physics and Astronomy, University of Glasgow, Glasgow G12 8QQ, United Kingdom.

To achieve this, we use the Green's function formalism and the $\hat{T}$-matrix equation to calculate the effect of a pointlike scatter

$$\hat{G}(\mathbf{R}, \mathbf{R}', \omega) = \hat{G}_0(\mathbf{R} - \mathbf{R}', \omega) + \hat{G}_0(\mathbf{R}, \omega)\hat{T}(\omega)\hat{G}_0(-\mathbf{R}', \omega), \quad (1)$$

where $\hat{G}_0(\mathbf{R}, \omega)$ is the noninteracting Green's function in discrete real space, with $\mathbf{R}$ describing a unit cell vector and $\omega$ the energy. Here, $\hat{G}_0(\mathbf{R}, \omega)$ is obtained from Fourier transform of the Green's function $\hat{G}_0(\mathbf{k}, \omega)$ in momentum space, a matrix in the basis of orbital and site positions $(s, p)$ whose elements are defined by

$$\hat{G}_0^{sp}(\mathbf{k}, \omega) = \sum_\mu \frac{\xi_\mu^{s\dagger}(\mathbf{k})\xi_\mu^p(\mathbf{k})}{\omega - \epsilon_\mu(\mathbf{k}) + i\Gamma}. \quad (2)$$

Here, $\epsilon_\mu(\mathbf{k})$ is the $\mu$th eigenvalue, and $\xi_\mu^s(\mathbf{k})$ the corresponding $s$th element of the $\mu$th eigenvector for the tight-binding Hamiltonian $H_0(\mathbf{k})$, which is described in the Supplemental Material [26]. $\Gamma$ is a small energy broadening, $\hat{T}(\omega) = \hat{V}[1 - \hat{V}\hat{G}^0(\mathbf{R} = 0, \omega)]^{-1}$ is the scattering matrix, and $\hat{V}$ is the scalar impurity potential. We then calculate the continuum Green's function [19,20]:

$$G(\mathbf{r}, \mathbf{r}', \omega) = \sum_{\mathbf{R}, \mathbf{R}', s, p} \hat{G}^{sp}(\mathbf{R}, \mathbf{R}', \omega) w_\mathbf{R}^s(\mathbf{r}) w_{\mathbf{R}'}^p(\mathbf{r}'), \quad (3)$$

where $\mathbf{r}$ describes the continuous real-space coordinate, and $w_\mathbf{R}^s(\mathbf{r})$ is the Wannier function of orbital $s$ on lattice site $\mathbf{R}$. The cLDOS, as measured by STM, can then be obtained via

$$\rho(\mathbf{r}, \omega) = -\frac{1}{\pi} \text{Im } G(\mathbf{r}, \mathbf{r}, \omega). \quad (4)$$

This method enables the detailed study of QPI, incorporating the wave function decay into the vacuum, as measured by STM, and has been shown to enable quantitative comparisons with experimental measurements [21–23,27–29].

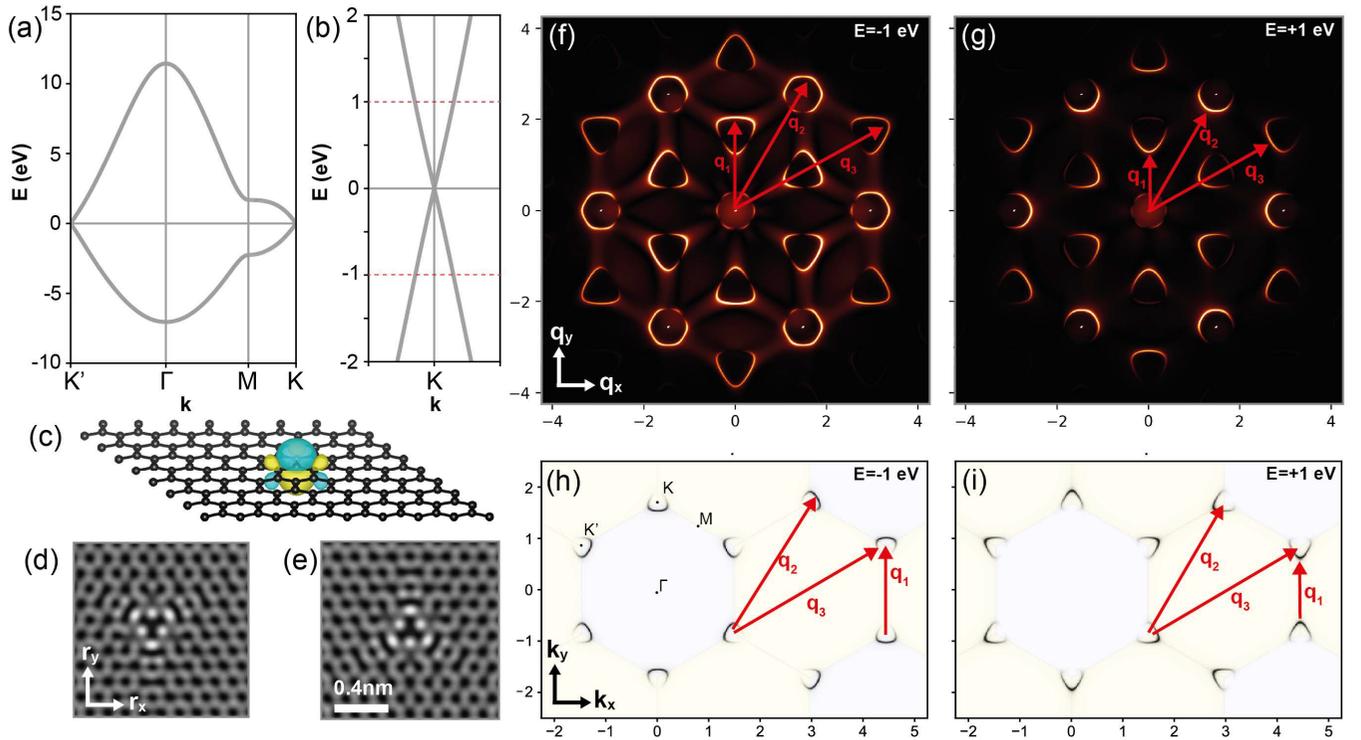

FIG. 1. Continuum quasiparticle interference (QPI) calculations for graphene. (a) Density functional theory (DFT)-derived band structure along $K'$-$\Gamma$-$M$-$K$ for the $p_z$ orbitals of graphene. (b) Close-up of the Dirac cone. Energies used in calculations in (f)–(i) are indicated by dashed horizontal lines. (c) Crystal structure of a monolayer of graphene, showing the Wannier function obtained from DFT used in the calculations. (d) and (e) Real-space continuum local density of states (cLDOS) calculations of graphene with a defect in either the A or B carbon atom for $E = +0.5$ eV. (f) and (g) Fourier transform of cLDOS calculations showing the QPI of graphene at (f) $E = -1$ eV and (g) $E = +1$ eV. (h) and (i) Unfolded spectral function of graphene at equivalent energies of (f) and (g), showing the importance of intraunit cell geometry, or pseudospin, in the physical observables of graphene. The red arrows in (h) and (i) can be directly traced back to prominent scattering vectors in (f) and (g). Both $\mathbf{k}$ and $\mathbf{q}$ have units of inverse angstroms.

Before discussing QPI in a moiré lattice, we first focus on the case of monolayer graphene. We employ a density functional theory (DFT)-derived tight-binding model of the $p_z$ orbitals of graphene to calculate the cLDOS, with the Wannier functions and model extracted using a modified version of WANNIER90 [30] to preserve the signs of the wave functions. The band structure obtained from the projection is shown in Figs. 1(a) and 1(b) and the wave functions in Fig. 1(c). QPI in graphene has been detected in experiments previously [31–34] and shows distinct energy-dependent semiarc shapes around the atomic Bragg peaks due to the pseudospin texture of the bands, which results in forbidden intra- or intervalley scattering, depending on whether one looks at energies above or below the Dirac point [33].

Figures 1(d) and 1(e) show the calculated real-space maps of the cLDOS centered around a defect placed on either the A or B carbon atom of the two-atom unit cell of graphene, respectively. The triangular shape of the scattering pattern around the defect is in excellent agreement with STM measurements [34], highlighting the fidelity of this method. By Fourier transforming the real-space cLDOS, we obtain the QPI scattering patterns in **q**-space shown in Figs. 1(f) and 1(g). We show QPI at representative energies $E = \pm 1$ eV, which show the semiarc QPI patterns that invert when changing energy across the Dirac point, in excellent agreement with experimental results [32,33,35].

Although it is informative to discuss these semiarc patterns in terms of the pseudospin texture of the Dirac cone in the honeycomb lattice, more generally, it is the spatial phase of the atoms within a unit cell that controls these selection rules. To highlight this, in Figs. 1(h) and 1(i), we plot the unfolded spectral function $A(\mathbf{k}, \omega)$ [24,25,36]. Different to the conventional spectral function, this quantity accounts for the intraunit-cell geometry via the positions of the atoms using

$$A(\mathbf{k}, \omega) = \sum_{s,p} \text{Im}\{G_0^{sp}(\mathbf{k}, \omega)\} \exp[i\mathbf{k}(\mathbf{r_s} - \mathbf{r_p})], \quad (5)$$

where **k** is the momentum, and $\mathbf{r}_i$ is the position of orbital $i$ within a unit cell. As shown in Figs. 1(g) and 1(h), the intensity of the pockets from the Dirac cones becomes strongly momentum and energy dependent, which is also seen and explained in the angle-resolved photoemission spectroscopy (ARPES) measurements for negative energies [37–43], with shapes that directly resemble the scattering patterns seen in the QPI pattern of Figs. 1(f) and 1(g). We indicate the three most prominent contributions to the scattering patterns by red arrows labeled $\mathbf{q}_1$, $\mathbf{q}_2$, and $\mathbf{q}_3$.

Having shown that the cLDOS approach can accurately reproduce the known QPI pattern of monolayer graphene and how it relates to the electronic structure by comparison with the unfolded spectral function, we now consider the QPI of TBLG. The moiré lattice of TBLG introduces a new unit cell the size of which depends on the twist angle and consists of many more atoms than the two found in the primitive unit cell of graphene. We model the electronic structure of TBLG using the local environment tight-binding model [44] and consider the effect of a defect on the QPI measured at the top layer closest to the STM tip; details can be found in the Supplemental Material [26]. We assume that the Wannier functions remain unchanged from the monolayer, which has been shown to be an accurate approximation for TBLG [45,46]. cLDOS calculations for TBLG with different twist angles reveal the real-space moiré pattern as shown in Fig. 2(a) for $\theta = 6.01°$ and $8.26°$. Formally, for $\theta = 8.26°$, the structure of TBLG has two distinct points of high intensity within a single unit cell where the atoms almost overlap, but not identically, known as a second-order moiré lattice [47], whereas for $\theta = 6.01°$, there is only one unique high-intensity point, known as a first-order Moiré lattice. Due to this difference, which is almost impossible to distinguish in the cLDOS, the unit cell can be much larger than one would be led to believe from simply considering the peak-to-peak distances from an STM topography. By looking at the cLDOS at a position far away from a defect [Fig. 2(b)], the VHs peak can be observed and is dependent on twist angle, in agreement with previous experiments [7]. Interestingly, despite significant changes to the unit cell and Brillouin zone for different twist angles, the appearance of a defect in the top layer does not show a significant change in shape, shown in Figs. 2(c)–2(f). However, the magnitude of the impurity bound state, discussed in the Supplemental Material [26], decreases with increasing twist angle, a trend also observed for small twist angles [48].

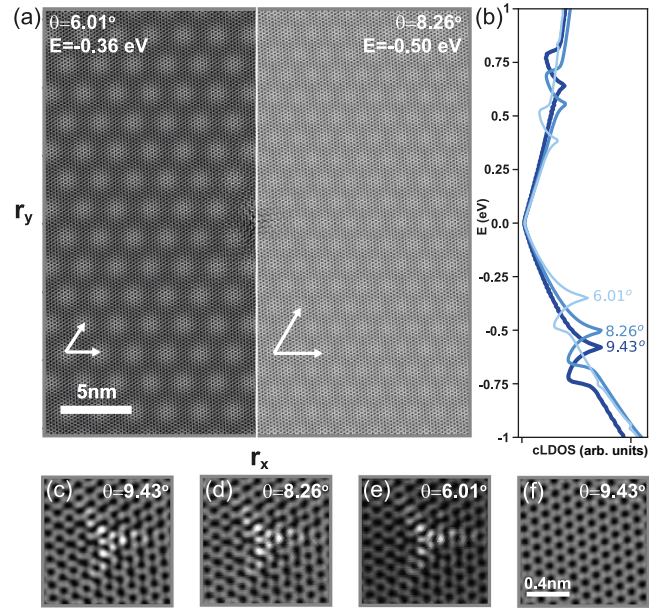

FIG. 2. Continuum local density of states (cLDOS) calculations of moiré patterns. (a) Real-space cLDOS $\rho(\mathbf{r}, E)$ for a first-order moiré angle $\theta = 6.01°$ and a second-order moiré angle $\theta = 8.26°$. The intensity of the moiré pattern is maximum at the energy of the Van Hove singularity (VHs), here, $E = 0.36$ and $0.50$ eV, respectively. (b) $\rho(\mathbf{r}, E)$ as a function of energy $E$ taken at a position **r** far away from the defect, highlighting the change of the VHs as a function of angle $\theta$. (c)–(e) Shape of a defect in the top layer and quasiparticle interference (QPI) in $\rho(\mathbf{r}, E)$ in real space for various twist angles $\theta$ at $E = +0.5$ eV. The shape of the defect and the surrounding QPI is nearly identical, with minimal influence from the bottom layer. (f) Real-space map of $\rho(\mathbf{r}, E)$ for a defect-free area with $\theta = 9.43°$ at $E = +0.5$ eV.

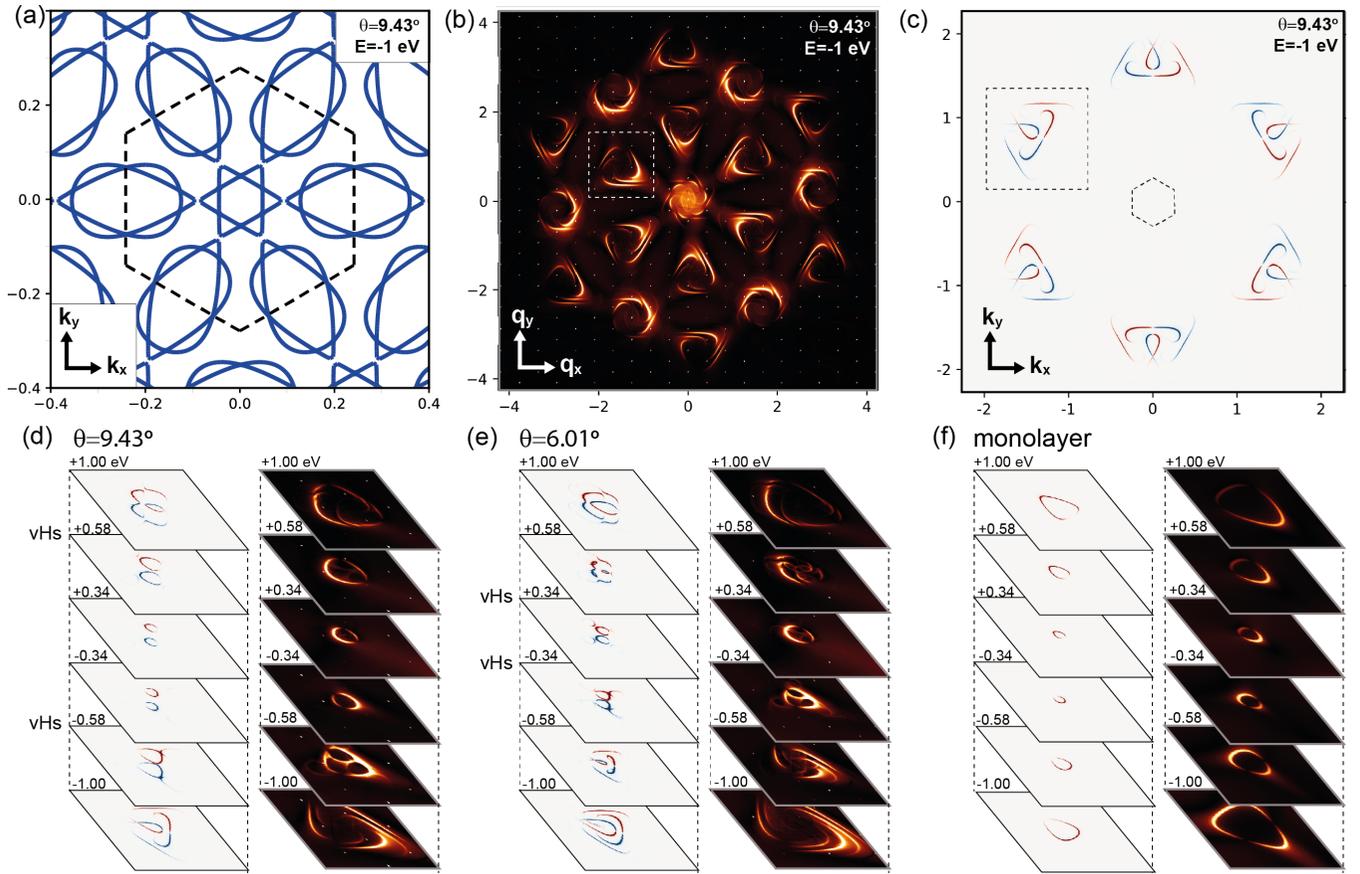

FIG. 3. Continuum quasiparticle interference (QPI) calculations of twisted bilayer graphene (TBLG). (a) Conventional constant energy contour of TBLG ($\theta = 9.43°$) at $E = -1$ eV. (b) QPI calculations at the same energy as (a), plotted on the same momentum scale as Figs. 1(h) and 1(i). (c) Unfolded spectral function $A(\mathbf{k}, \omega)$ of the same model as in (a), the dashed hexagonal shows the first Brillouin zone. (d)–(f) Constant energy cuts of the main features shown in the unfolded spectral function (left) and QPI (right) shown by the dashed boxes in (b) and (c) for (d) $\theta = 9.43°$ (e) $\theta = 6.01°$, and (f) monolayer graphene. Here, the red and blue colors in the unfolded spectral function plots highlight the projected weight from the top and bottom layers of graphene, respectively.

The presence of the moiré lattice greatly expands the unit cell of TBLG and conversely reduces the size of the Brillouin zone, inducing backfolding effects, as shown in the constant energy contour of the electronic structure in Fig. 3(a). This would lead one to believe that the QPI pattern should also be dramatically modified; however, we show here that this is not the case. In Fig. 3(b), we present the Fourier transform of the cLDOS for $\theta = 9.43°$ on the same **q**-scale as for monolayer graphene in Figs. 1(f) and 1(g). Although many more Bragg peaks corresponding to the periodicity of the moiré lattice can now be observed, the dominant QPI scattering vectors remain similar to those of monolayer graphene. To understand this, we compare the calculations with the unfolded spectral function, which now incorporates hundreds of atoms within a unit cell. The intraunit-cell phase factors between the atoms destructively interfere everywhere except around the $K$-points of the Brillouin zone of the individual graphene sheets, as shown in Fig. 3(c). The two valleys at the $K$ points are rotated by $\theta$, leading to two Dirac cones, one from the top monolayer (blue) and the other from the bottom monolayer (red), that overlap and hybridize to form a distinct pattern, similar to that observed in ARPES [49–52]. As we have included a pointlike defect in our calculation in the top-layer only (calculations for the bottom layer may be found in the Supplemental Material [26]), the defect interacts much more strongly with the spectral weight from the top monolayer, leading to strong QPI signals at similar momentum to monolayer graphene, with similar cone shapes that are partially gapped and directly match the shape of either the blue or red cones in Fig. 3(c). This similarity to monolayer graphene becomes even more pronounced at energies closer to the Dirac point, as shown in Fig. 3(d). In the spectral function, the VHs emerges at the energy ($E_{\text{VHs}}$) where the Dirac cones of the two graphene layers touch, which produces a strong hybridization pattern in the QPI. At energies $|E| < |E_{\text{VHs}}|$, the QPI is almost identical to a single Dirac cone of monolayer graphene [compare Fig. 3(f)]. Changing the twist angle from $\theta = 9.43°$ to $6.03°$ [Fig. 3(e)] confirms that this behavior holds for other twist angles. This is a directly testable signature of the VHs induced by the backfolding of the moiré lattice in QPI.

In this letter, we provide a direct method and understanding to quantitatively compare model $E(k)$ electronic structures with QPI, unlocking the ability to extract the experimental electronic structure of twisted bilayer materials from STM measurements. While a naive interpretation would suggest that gigantic fields of view covering several moiré unit cells are required to resolve the features close to $\mathbf{q} = 0$, our results show that the QPI signal close to the atomic peaks contains the same information. Large fields of view are still required for sufficient $\mathbf{q}$-resolution but without the need to cover multiple moiré unit cells. We also show that considering the unfolded spectral function helps with understanding key features of the QPI of the twisted heterostructure, which will have significance in the interpretation of further moiré phenomena, e.g., chiral Bogoljubov quasiparticles [12].

In this letter, we focused on identifying the connection between moiré lattices and QPI at relatively large twist angles that neglect the effect of lattice relaxations and electron-electron interactions. Nevertheless, these calculations provide a starting point to assess the impact of these effects by enabling direct comparison of the electronic structure as measured by QPI with experimental data. Small twist angles can be studied by using continuum models and Wannier functions [14] using the framework described above, as can lattice relaxations by modification of the initial hopping parameters. We have also previously shown that the sign structure of the superconducting order parameter can be predicted within this framework [53], which can therefore provide a direct test of the nature of superconductivity in TBLG and thus crucial information about the potential coupling mechanism.

We thank Peter Hirschfeld for useful discussions. L.C.R. and P.W. acknowledge funding from the Leverhulme Trust through Research Project Grant No. RPG-2022-315. We used computational resources of the Cirrus UK National Tier-2 HPC Service at EPCC [54] funded by the University of Edinburgh and EPSRC (No. EP/P020267/1). P.W. gratefully acknowledges the access to and use of computational resources of the HPC cluster Marvin hosted by the University of Bonn.